\documentclass[pdflatex,sn-mathphys-num]{sn-jnl}

\usepackage{graphicx}%
\usepackage{multirow}%
\usepackage{amsmath,amssymb,amsfonts}%
\usepackage{amsthm}%
\usepackage{mathrsfs}%
\usepackage[title]{appendix}%
\usepackage{xcolor}%
\usepackage{textcomp}%
\usepackage{manyfoot}%
\usepackage{booktabs}%
\usepackage{algorithm}%
\usepackage{algorithmicx}%
\usepackage{algpseudocode}%
\usepackage{listings}%
\usepackage{booktabs}

\usepackage{caption}

\theoremstyle{thmstyleone}%
\usepackage{lineno}
\theoremstyle{thmstyletwo}%
\theoremstyle{thmstylethree}%
\begin{document}

\title{Self-motion as a structural prior for coherent and robust formation of cognitive maps}

\author[1,2]{\fnm{Yingchao} \sur{Yu}} \email{yingchaoyu@mail.dhu.edu.cn}\equalcont{These authors contributed equally to this work.}
\author[3]{\fnm{Pengfei} \sur{Sun}} \email{p.sun@imperial.ac.uk}\equalcont{These authors contributed equally to this work.}
\author*[2]{\fnm{Yaochu} \sur{Jin}}\email{jinyaochu@westlake.edu.cn}
\author[1]{\fnm{Kuangrong} \sur{Hao}}\email{krhao@dhu.edu.cn}
\author[2]{\fnm{Hao} \sur{Zhang}}\email{haozhang461000@gmail.com}
\author[4]{\fnm{Yifeng} \sur{Zhang}}\email{peter.zyf@alibaba-inc.com}
\author[2]{\fnm{Wenxuan} \sur{Pan}}\email{panwenxuan@westlake.edu.cn}
\author[2]{\fnm{Wei} \sur{Chen}}\email{chenwei06@westlake.edu.cn}
\author[3,6,7]{\fnm{Danyal} \sur{Akarca}}\email{d.akarca@imperial.ac.uk}
 
\author[5]{\fnm{Yuchen} \sur{Xiao}}\email{xiaoyuchen@westlake.edu.cn}

\affil[1]{\orgdiv{School of Information and Intelligence Science}, 
          \orgname{Donghua University}, 
          \orgaddress{\city{Shanghai}, 
          \country{China}}}

\affil[2]{\orgdiv{School of Engineering}, 
          \orgname{Westlake University}, 
          \orgaddress{\city{Hangzhou}, 
          \country{China}}}

\affil[3]{\orgdiv{Department of Electrical and Electronic Engineering},            \orgname{Imperial College London},
          \orgaddress{\city{London}, 
          \country{UK}}}

\affil[4]{\orgname{Alibaba Group}, 
          \orgaddress{\city{Hangzhou}, 
          \country{China}}}

\affil[5]{\orgdiv{School of Life Sciences}, 
          \orgname{Westlake University}, 
          \orgaddress{\city{Hangzhou}, 
          \country{China}}}          

\affil[6]{\orgdiv{Imperial-X}, 
          \orgname{Imperial College London},          \orgaddress{\city{London}, 
          \country{UK}}}

\affil[7]{\orgdiv{MRC Cognition and Brain Sciences Unit}, 
          \orgname{University of Cambridge},
          \orgaddress{\city{Cambridge}, 
          \country{UK}}}
          
\abstract{
Most computational accounts of cognitive maps assume that stability is achieved primarily through sensory anchoring, with self-motion contributing to incremental positional updates only. However, biological spatial representations often remain coherent even when sensory cues degrade or conflict, suggesting that self-motion may play a deeper organizational role. Here, we show that self-motion can act as a structural prior that actively organizes the geometry of learned cognitive maps. We embed a path-integration-based motion prior in a predictive-coding framework, implemented using a capacity-efficient, brain-inspired recurrent mechanism combining spiking dynamics, analog modulation and adaptive thresholds.  
Across highly aliased, dynamically changing and naturalistic environments, this structural prior consistently stabilizes map formation, improving local topological fidelity, global positional accuracy and next-step prediction under sensory ambiguity. 
Mechanistic analyses reveal that the motion prior itself encodes geometrically precise trajectories under tight constraints of internal states and generalizes zero-shot to unseen environments, outperforming simpler motion-based constraints. Finally, deployment on a quadrupedal robot demonstrates that motion-derived structural priors enhance online landmark-based navigation under real-world sensory variability.
Together, these results reframe self-motion as an organizing scaffold for coherent spatial representations, showing how brain-inspired principles can systematically strengthen spatial intelligence in embodied artificial agents.
}
\maketitle

\section{Introduction}\label{sec1}
Animals can maintain continuously updated internal representations of their surroundings, known as cognitive maps \cite{tolman1946studies, mou2025representing, whittington2022build, behrens2018cognitive, yu2025structural}, despite the often noisy, ambiguous and continually fluctuating sensory inputs that they rely on. A prevailing assumption across neuroscience and computational modeling is that the stability of these maps arises primarily on sensory anchoring, with self-motion signals serving only incremental updates to an underlying position \cite{kessler2024human, naveilhan2025theta, knierim1995place, gothard1996dynamics, mao2025multisensory, savelli2019origin, jayakumar2019recalibration}. This view is embedded in many existing models, which treat self-motion as moment-to-moment displacement information \cite{banino2018vector, whittington2020tolman, kessler2024human} rather than as a contributor to the formation or global coherence of the spatial map itself. However, this sensory-anchoring-centric view struggles to account for a recurring empirical observation: biological spatial representations often remain coherent even when sensory cues are unreliable, ambiguous, or even contradictory \cite{wen2024one, knierim1998interactions, yartsev2013representation}, suggesting that self-motion may play a role beyond simple positional updating.

This limitation becomes particularly evident in a range of biological experiments. Spatial maps frequently remain coherent, or reorganize in systematic, non-random ways, even when sensory cues are degraded or conflicting. In darkness, grid-cell activity drifts coherently while preserving its lattice structure \cite{wen2024one, fyhn2007hippocampal}; when self-motion cues conflict with external landmarks, maps systematically realign rather than collapse \cite{knierim1998interactions, jeffery1999learned}; and when animals explore along restricted or biased trajectories, the map geometry systematically deforms in ways that reflect their movement statistics \cite{yartsev2013representation, mehta1997experience}. Although these phenomena are often interpreted as minor forms of path-integration drift or circuit-specific quirks, they share a common computational signature: when sensory cues underspecify the environment, self-motion-related dynamics continue to impose coherent structure on the evolving spatial representation. 

\begin{figure}[!htbp]
\centering
\includegraphics[width=0.95\linewidth]{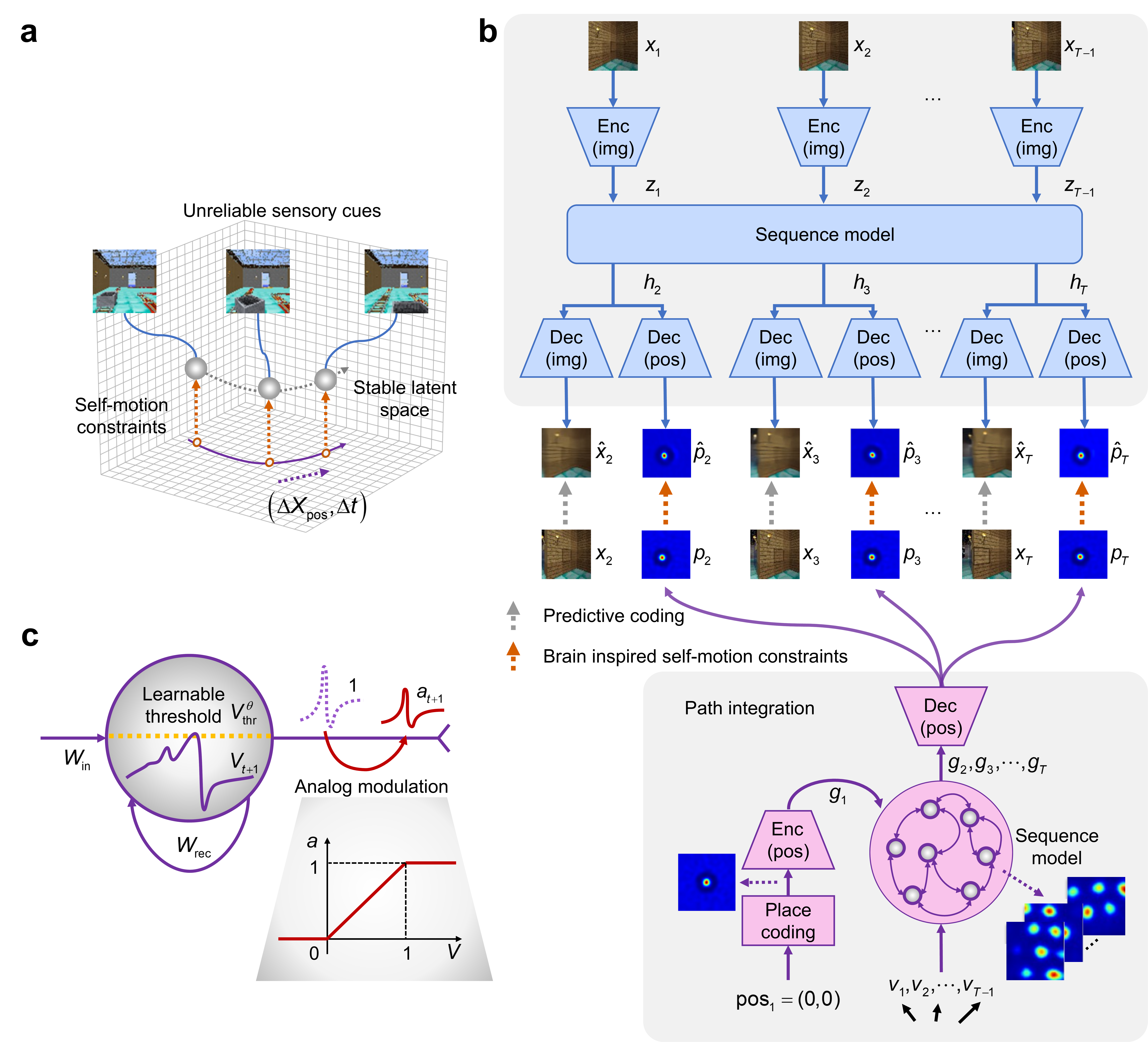}
\caption{
\textbf{A brain-inspired path-integration prior for coherent cognitive-map formation.}
\textbf{a}  Conceptual illustration of the hypothesized role of self-motion as a geometric and temporal scaffold that constrains the evolution of latent spatial representations when sensory cues are unreliable or ambiguous.
\textbf{b}  Self-motion is implemented as a path-integration (PI) mechanism and incorporated as a representation-level structural prior within the predictive-coding framework. 
\textbf{c}  The brain-inspired PI neuron integrates recurrent spiking dynamics, analog membrane modulation, and adaptive firing thresholds to support stable, geometrically accurate, and capacity-efficient integration.
}
\label{img1}
\end{figure}

This pattern motivates a reinterpretation of the role of self-motion in spatial cognition. Rather than serving solely as an incremental update signal between successive sensory anchors, self-motion may provide a geometric and temporal scaffold that constrains how spatial representations can evolve over time. Under this view, coherence arises not only from external reference cues, but from internally consistent trajectory dynamics that organize the latent space, particularly when sensory evidence is sparse or ambiguous (\textbf{Figure \ref{img1}a}). If self-motion indeed functions as a structural prior, then incorporating motion-consistent constraints into representation learning should yield specific, testable consequences. In particular, such constraints should stabilize learned cognitive maps under sensory ambiguity, preserve local topological structure, strengthen global positional information, and improve predictive continuity, even when sensory inputs alone fail to uniquely specify location.

To evaluate this hypothesis, we integrate self-motion constraints into a predictive-coding (PC) framework for autonomous cognitive-map construction (\textbf{Figure~\ref{img1}b}). Crucially, self-motion is not treated as an auxiliary input or as an explicit supervision signal, rather as a representation-level constraint that shapes the evolution of latent states during learning. To obtain a more precise yet capacity-efficient source of self-motion information, we instantiate this constraint using a brain-inspired path-integration (PI) mechanism whose recurrent spiking dynamics incorporate analog modulation and adaptive thresholding, features observed in CA3 neurons (\textbf{Figure~\ref{img1}c}) \cite{alle2006combined, trinh2023adaptive, sasaki2011action, zbili2019past}. The recurrent architecture of CA3 is known to support stable, coherent, and high-fidelity reinstatement of spatial activity patterns within a relatively compact neuronal population \cite{watson2025human, neunuebel2014ca3, mizuseki2012activity, west1990unbiased}, making it an organizational template well suited for capacity-limited implementations of path integration. This biological circuitry therefore offers an existence proof that robust, high-fidelity path integration can be realised within compact and noisy systems, supplying design principles that are typically absent from conventional engineering-based PI modules.

Across our experiments, we show that incorporating self-motion as a structural constraint markedly stabilizes cognitive-map formation, improving local topological preservation, strengthening global positional information and enhancing next-step predictive accuracy under severe perceptual aliasing, dynamically changing environments and naturalistic sensory variability. We further demonstrate that a brain-inspired path-integration mechanism provides a clear mechanistic advantage, supplying a fast-converging, geometrically accurate, and neurally consistent motion prior whose benefits persist at high spatial resolution, under tight internal-state capacity constraints and during zero-shot generalization in unseen environments. Importantly, the underlying principle is general: multiple forms of self-motion constraint, including simple velocity signals, improve map stability, while the brain-inspired mechanism yields the strongest and most reliable gains. Extending these findings to an embodied setting, deployment on a quadrupedal robot confirms that motion-based constraints remain robust under real-world sensory noise and physical environmental variability.

Together, our results show that self-motion provides a structural scaffold for cognitive maps, stabilizing their organization beyond what sensory cues alone can support. This insight reframes the way both biological and artificial systems can construct reliable representations under conditions of uncertainty. More broadly, the findings suggest that internally generated dynamics, which have long been treated as a potential source of drift, may instead act as a fundamental structural prior, shaping how both brain and machines learn coherent representations when the external world underspecifies them.

\section{Results}\label{sec2}
\subsection{Brain inspired self-motion constraints stabilize cognitive-map formation}

We investigate the formative role of self-motion in construction of cognitive maps by embedding a path integration mechanism in a predictive-coding framework (\textbf{Figure~\ref{img1}b}). The PI module functions as an internally generated geometric prior: it supplies the predictive-coding system with a world-consistent estimate of how latent spatial structure should evolve under self-motion. Because this prior is computed independently of visual input, it provides a source of stable geometric organisation even when appearance cues are ambiguous or rapidly changing. 

To obtain a more capacity-efficient yet finer-grained form of this prior, we implement a brain-inspired PI mechanism motivated by the CA3 circuit (a subregion of the hippocampus \cite{li1994hippocampal, treves1994computational}). Although CA3 is not believed to perform path integration directly, its recurrent architecture supports stable, high-fidelity spatial activity patterns within a relatively compact neuronal population \cite{neunuebel2014ca3, mizuseki2012activity, west1990unbiased}, matching the requirements we place on PI under limited model capacity. We therefore capture key CA3-like properties by implementing recurrent spiking dynamics with analog modulation and learnable thresholds \cite{alle2006combined, trinh2023adaptive, sasaki2011action, zbili2019past} (\textbf{Figure~\ref{img1}c}). This module generates a temporally coherent latent trajectory that reflects an underlying geometry of self-motion, and the predictive-coding model learns to align its latent state with this trajectory during training, thereby linking sensory prediction to an internally generated motion structure.

\begin{figure}[t!]
\centering
\includegraphics[width=0.98\linewidth]{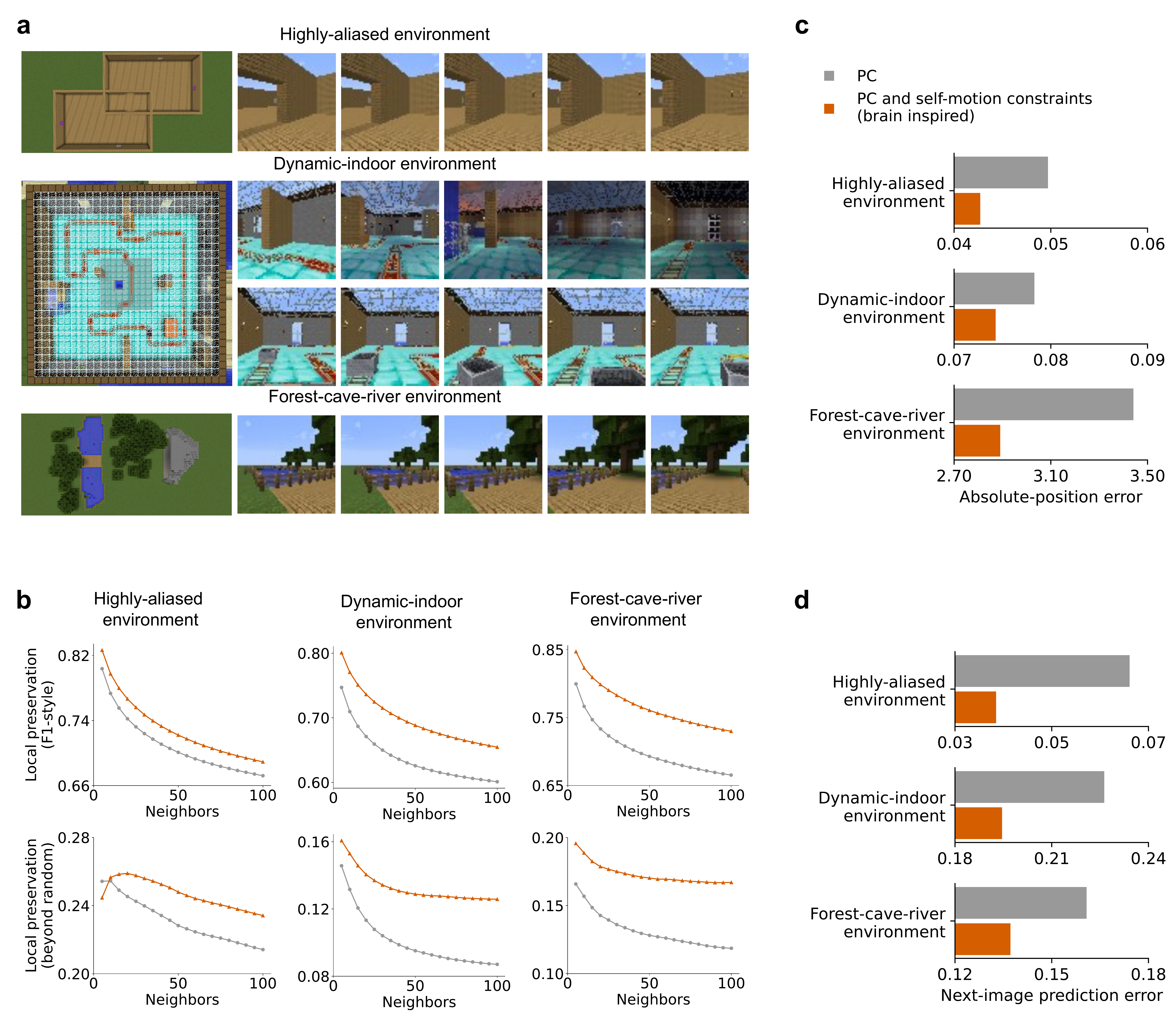}
\caption{
\textbf{Brain-inspired self-motion priors improve cognitive-map formation across diverse environments.}
\textbf{a}  Three simulated environments used for evaluation: a highly aliased environment and a dynamic indoor environment designed in this work, and a naturalistic forest-cave-river benchmark \cite{gornet2024automated}.  
\textbf{b}  Self-motion priors consistently enhance local neighborhood structure, reflected in higher $\mathrm{H}(k)$ (F1-like measures of neighborhood preservation) and $\mathrm{LCMC}(k)$ (local continuity beyond random chance) scores.
\textbf{c}  Models trained with the self-motion prior encode more accurate global positional information, as assessed by a downstream predictor.   
\textbf{d}  Next-image prediction error decreases across all environments, indicating stronger predictive continuity under the constraint.
}
\label{res1}
\end{figure}

We evaluate this framework in three challenging simulated environments designed to stress different aspects of spatial inference: a highly aliased environment, a dynamic indoor environment, and a forest-cave-river environment (\textbf{Figure~\ref{res1}a}). The first two environments are designed to isolate the effects of self-motion constraints when visual cues are ambiguous or rapidly changing, whereas the forest-cave-river environment serves as a publicly available benchmark with more naturalistic sensory variability \cite{gornet2024automated}.

If self-motion acts as a structural prior, its most immediate effect should be the preservation of local spatial neighborhoods in the latent map.  Incorporating brain-inspired self-motion constraints substantially improves the preservation of local neighborhood structure in the learned cognitive maps (\textbf{Figure~\ref{res1}b}). Local structure is quantified using the harmonic mean of Trustworthiness and Continuity ($\mathrm{H}(k)$), together with the local-continuity meta-criterion ($\mathrm{LCMC}(k)$) across neighborhood sizes $k$ (see \textbf{Methods}). Despite differences between environments, the benefits remain consistent. 
In the highly-aliased environment, the average $\mathrm{H}(k)$ across all neighborhood sizes $k$ increased by 2.9\%, and $\mathrm{LCMC}(k)$ by 6.9\%. 
The gains were greater in the dynamic-indoor environment, reaching 9.4\% and 32.3\%, respectively, and remained strong in the forest-cave-river environment (9.3\% and 31.5\%). 
Notably, at very small neighborhoods (e.g. $k=5$), $\mathrm{LCMC}$ can decrease slightly because the constraints can suppress spurious appearance-based micro-clusters, trading nearest-neighbor continuity for more spatially meaningful structure.

We further assess the global positional information encoded in the latent cognitive maps. Across all environments, incorporating self-motion constraints consistently improved global positional accuracy (\textbf{Figure~\ref{res1}c}), reducing error by 14.1\%, 5.1\% and 16.0\% in the highly-aliased, dynamic-indoor, and forest-cave-river environments, respectively. A similar pattern emerged for anticipatory sensory inference: next-image prediction errors decreased substantially across settings (\textbf{Figure~\ref{res1}d}), including a 41.8\% reduction under strong aliasing. 

Together, these results show that the brain inspired self-motion constraints provide a robust organizing signal for cognitive maps, consistently strengthening different aspects of the representation as sensory information becomes unreliable.

\subsection{Complementary brain-inspired properties underpin an accurate and neurally consistent self-motion prior}

The observation that brain-inspired self-motion constraints substantially stabilize cognitive-map formation prompts a natural mechanistic question: What specific advantage does the brain-inspired PI offer over conventional PI? This question is particularly compelling because conventional PI has rarely been used as an explicit constraint, and its capacity to function as a structural prior remains largely unexplored.

\begin{figure}[t]
\centering
\includegraphics[width=\linewidth]{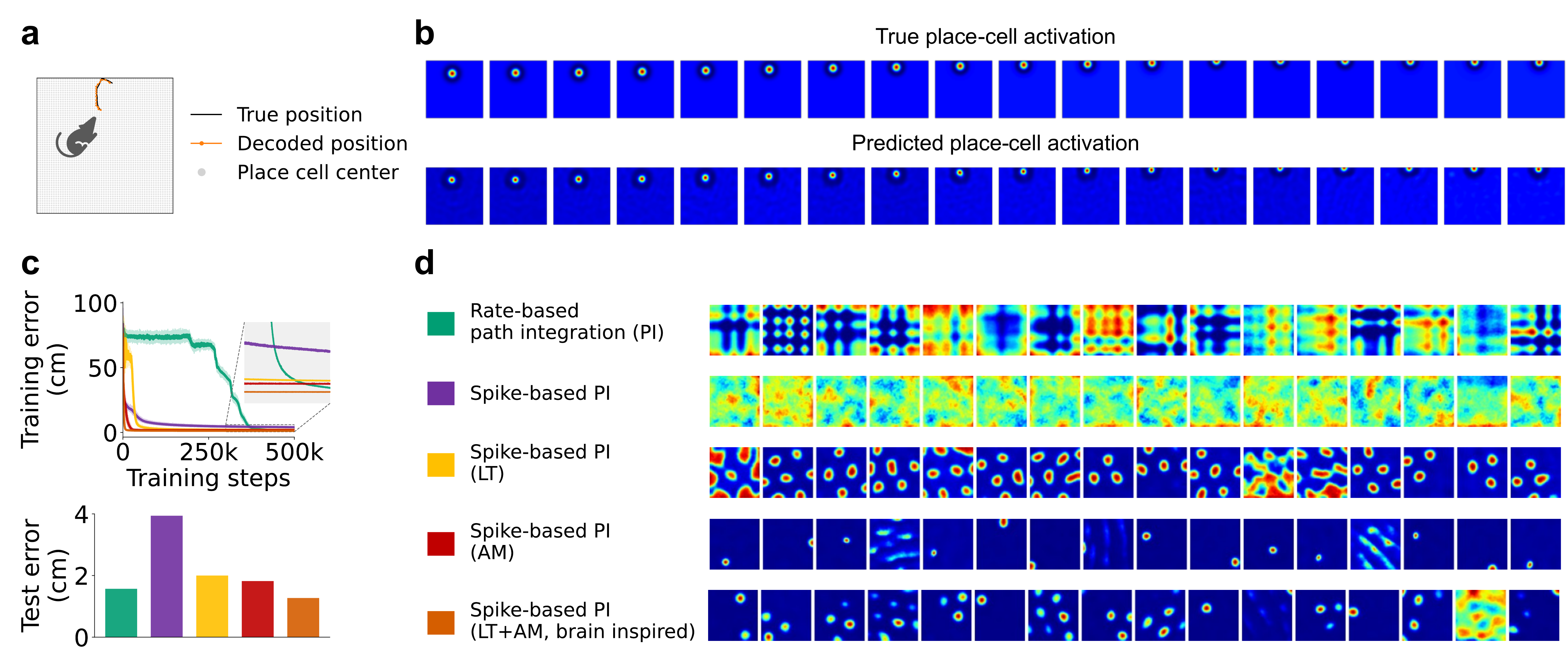}
\caption{
\textbf{Complementary CA3-derived properties shape an accurate and neurally consistent self-motion prior.}
\textbf{a}  A controlled place-cell decoding task constructed from simulated rodent trajectories \cite{raudies2012modeling, banino2018vector}.   
\textbf{b}  Ground-truth place-cell activity and corresponding predictions produced by the PI module.
\textbf{c}  Test error and training dynamics for five PI variants, revealing substantial differences in convergence speed and integration accuracy.  
\textbf{d}  Spatial coding patterns learned by each variant, showing that only variants incorporating both spiking dynamics and adaptive thresholds form coherent grid-like structure.
\textbf{Abbreviations:} LT, learnable thresholds; AM, analog modulation.
}
\label{res2}
\end{figure}

\begin{figure}[!htbp]
\centering
\includegraphics[width=\linewidth]{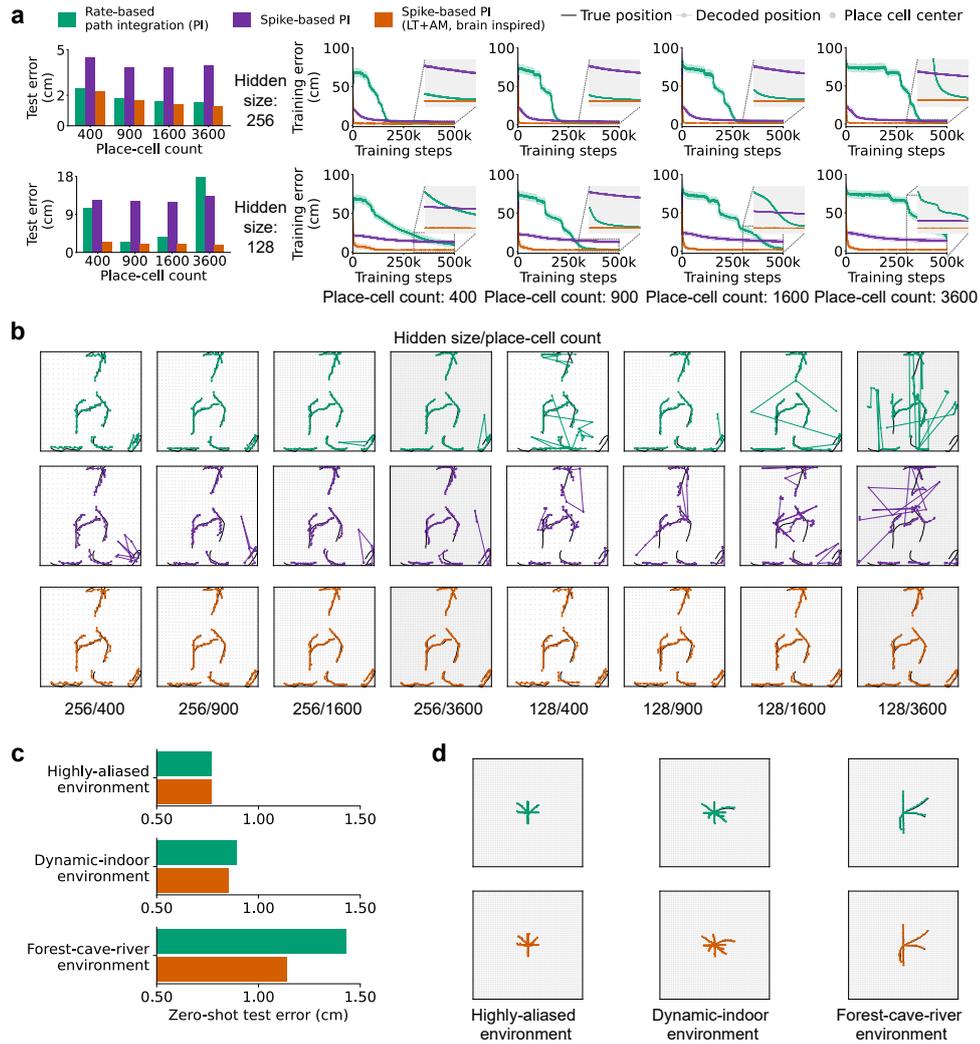}
\caption{
\textbf{Brain-inspired path integration remains accurate and stable across varying spatial resolutions, capacity limits and environments.}
\textbf{a}  Path integration error for rate-based, spike-based and brain-inspired variants across multiple place-cell resolutions and hidden-state capacities, revealing a growing advantage of the brain-inspired mechanism as representational capacity becomes more constrained.  
\textbf{b}  Representative reconstructed trajectories across configurations, demonstrating reduced drift and higher geometric fidelity for the brain-inspired variant.  
\textbf{c}  Zero-shot path-integration accuracy of rate-based and brain-inspired PI across three environments.  
\textbf{d}  Corresponding zero-shot trajectory reconstructions, showing more accurate and temporally stable paths produced by the brain-inspired mechanism.
}
\label{res22}
\end{figure}

Rather than attributing the advantage to a single design choice, we ask whether complementary properties jointly contribute to the observed stabilization. We hypothesize that the advantage of the brain-inspired PI arises from three complementary properties, motivated by CA3 physiology: (1) recurrent spiking dynamics that suppress small fluctuations through the discreteness of spikes, stabilizing integration trajectories and enabling faster convergence; (2) analog modulation that supports finer-grained accumulation of self-motion signals under limited capacity; and (3) learnable thresholds that enhance salient motion cues while suppressing noise. Together, these properties should yield a faster-converging, more geometrically faithful, and more capacity-efficient self-motion prior than conventional PI mechanisms.

We disentangle the contribution of each property by constructing PI variants with different combinations of spike-based recurrence, analog modulation, and learnable thresholds. We assess these variants in a controlled place-cell-decoding task using simulated rodent trajectories \cite{raudies2012modeling, banino2018vector} (\textbf{Figure~\ref{res2}a,b}). Because the PI module receives only low-dimensional velocity inputs and must reconstruct a high-dimensional place-cell representation using its limited internal state, the task imposes strict capacity constraints that make it well suited for isolating mechanistic contributions. We begin with a challenging configuration of 3600 place cells and a hidden size of 256, comparing five variants: a rate-based PI with continuous recurrent integration, a spike-based PI, a spike-based PI with a learnable threshold, a spike-based PI with analog modulation, and the full brain-inspired PI.

We find that the rate-based PI consistently converges substantially more slowly than all spike-based variants (\textbf{Figure~\ref{res2}c, upper panel}), consistent with the hypothesis that the discreteness of spikes suppresses small fluctuations, thereby accelerating convergence. Adding learnable thresholds or analog modulation further improved accuracy: test errors decreased by 49.2\% with learnable thresholds and by 53.8\% with analog modulation (\textbf{Figure~\ref{res2}c, lower panel}). Notably, the full brain-inspired PI achieved both the lowest path-integration error and the fastest convergence among all variants, reducing test error by 67.8\% relative to the spike-based PI and by 19.1\% relative to the rate-based PI.

Beyond integration accuracy, the PI variants exhibit distinct spatial coding patterns (\textbf{Figure~\ref{res2}d}). The brain-inspired PI and the spike-based PI with a learnable threshold develop clear grid-like responses, whereas the spike-based PI and the spike-based PI with analog modulation fail to form structured grids. In contrast, the rate-based PI instead produces square grid-like patterns rather than the hexagonal codes typical of biological systems, even under non-negative activation constraints \cite{dordek2016extracting, sorscher2023unified}; under our capacity-limited setting, this square symmetry may reflect restricted representational degrees of freedom. These observations indicate that adaptive thresholds are critical for hexagonal grid formation and that spiking dynamics potentially supports more biologically consistent spatial codes.

Together, these results confirm that the three properties inspired by CA3 provide complementary benefits, jointly yielding a faster-converging, more accurate, and more biologically consistent self-motion prior.

\subsection{Brain-inspired PI exhibits robust and scalable generalization across capacities and environments}

Building on the mechanistic insights above, we first assess the advantages of the brain-inspired PI across a wider range of model capacities by comparing it with the spike-based and rate-based PI under two tightly constrained hidden sizes (256 and 128) and four spatial resolutions (400, 900, 1600, and 3600 place cells). Across all configurations, the brain-inspired PI consistently achieves the lowest test error. Its advantage becomes increasingly pronounced under tight capacity: with only 128 hidden units and 3600 place cells, test error decreased by 86.6\% relative to the spike-based PI and by 90.0\% relative to the rate-based PI. Although the rate-based PI performs moderately well at intermediate resolution, it degrades rapidly at higher resolutions,  reflecting the fragility of continuous-valued integration under stringent capacity limits (\textbf{Figure~\ref{res22}a, left panel}).

\begin{figure}[!htbp]
\centering
\includegraphics[width=0.94\linewidth]{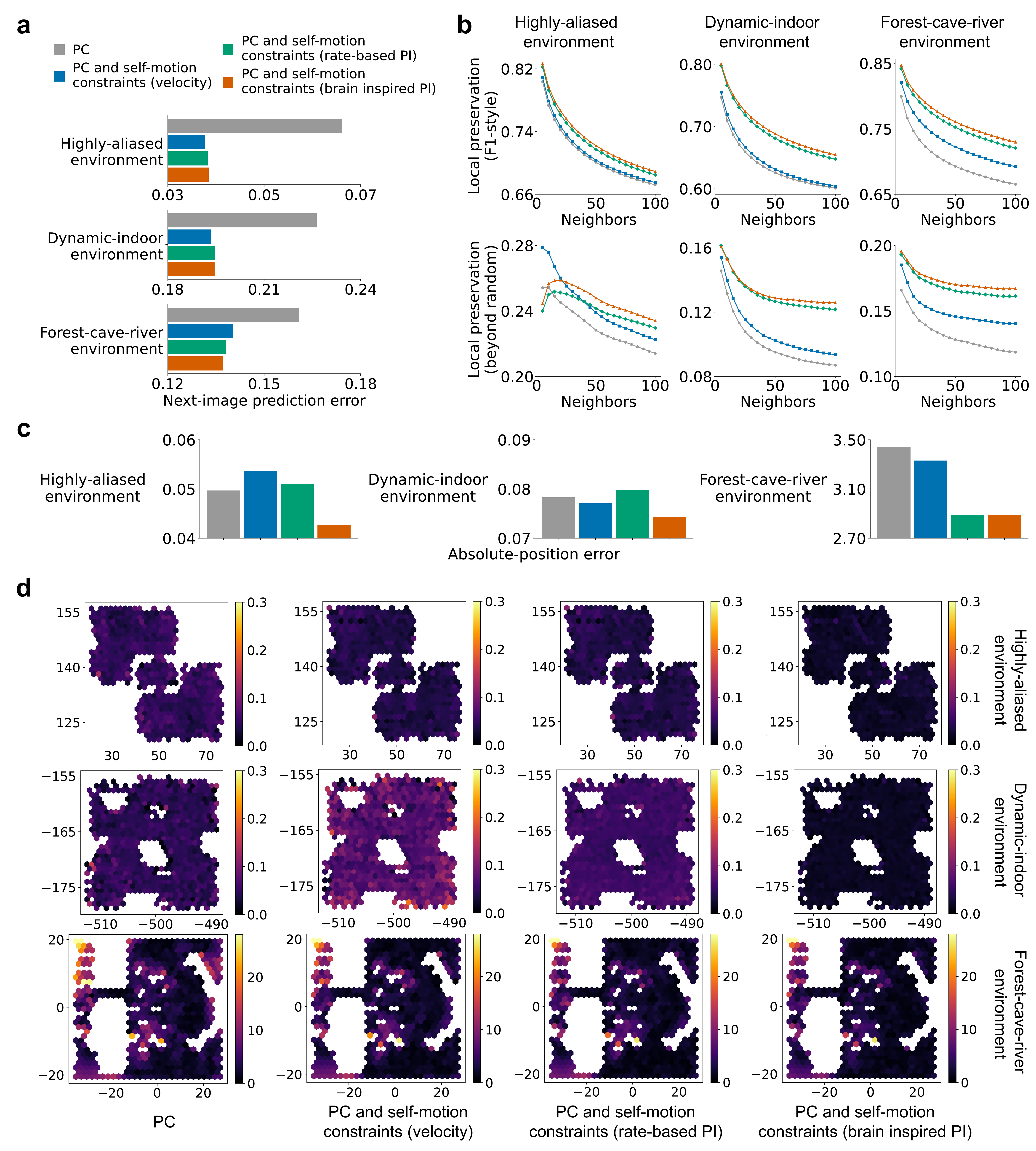}
\caption{
\textbf{Self-motion constraints act as a general scaffold for map construction, with the brain-inspired prior yielding the most consistent benefits.}
\textbf{a}  Next-image prediction error across environments for three forms of self-motion constraint: velocity inputs, rate-based PI, and brain-inspired PI, showing that all constraints improve predictive continuity.
\textbf{b}  Local neighborhood preservation measured by the harmonic mean of Trustworthiness and Continuity ($\mathrm{H}(k)$) and the local-continuity meta-criterion ($\mathrm{LCMC}(k)$), indicating larger gains for PI-based constraints than velocity alone, with the brain-inspired PI providing the strongest improvements.
\textbf{c}  Absolute-position estimation reveals clearer dissociations: velocity and rate-based PI improve performance only in selected environments, whereas the brain-inspired PI consistently reduces positional error across all settings.
\textbf{d}  Heatmaps of absolute-position prediction error (darker colors indicate lower error), showing that the brain-inspired constraint produces the most spatially coherent and robust global position estimates.
}
\label{res3}
\end{figure}

Training dynamics reveal parallel trends. The brain-inspired PI converges substantially faster, rapidly entering low-error regimes by 50k steps, whereas the rate-based PI requires 500k steps to reach comparable error levels and still produces only blurry square patterns (\textbf{Supplementary Figure~\ref{sup1}a-c}). As capacity tightens, the rate-based PI becomes increasingly unstable, the spike-based PI improves but remains less efficient, and the brain-inspired PI maintains rapid convergence (\textbf{Figure~\ref{res22}a, right panel}). Reconstructed trajectories further reinforce these distinctions (\textbf{Figure~\ref{res22}b}): the brain-inspired PI produces smooth, coherent, and geometrically accurate paths across all configurations, closely matching ground truth even in the most capacity-limited case (128 units, 3600 cells), whereas the other variants drift markedly from the true path, with errors sharply amplified under high-resolution and low-capacity conditions. 

After establishing that the brain-inspired PI performs robustly under varying capacity and resolution constraints, we next examine whether such robustness carries over to new environments. We therefore evaluate the brain-inspired PI in a zero-shot manner across all three environments, with models pretrained in the controlled place-cell-decoding task \cite{raudies2012modeling, banino2018vector}. We use the configuration where both the brain-inspired PI and the rate-based PI perform best during pretraining (3600  cells, 256 units).  In this zero-shot setting, both models achieve low path-integration error, but the brain-inspired PI performs consistently slightly better (\textbf{Figure~\ref{res22}c}) and yields trajectory reconstructions that more closely match the ground truth (\textbf{Figure~\ref{res22}d}).

We further probe the sensitivity by varying the latent-state integration factor $\alpha$. Path-integration error remains low across a broad range of $\alpha$ values, indicating that the {brain-inspired PI}
remains robust under broad operating conditions (\textbf{Supplementary Figure~\ref{sup2}a}). 
The corresponding spatial coding patterns vary smoothly with $\alpha$, transitioning from grid-like to place-like representations without abrupt regime changes as $\alpha$ increases (\textbf{Supplementary Figure~\ref{sup2}b}). This smooth dependence is incompatible with accidental parameter tuning and instead reflects a continuous deformation of a single underlying dynamical mechanism.

Together, these results show that the CA3-inspired mechanisms confer a robust and scalable advantage in path integration across capacities, resolutions, environments, and hyperparameters, underscoring the reliability and broad applicability of its structurally grounded motion prior.

\begin{figure}[!htbp]
\centering
\includegraphics[width=0.92\linewidth]{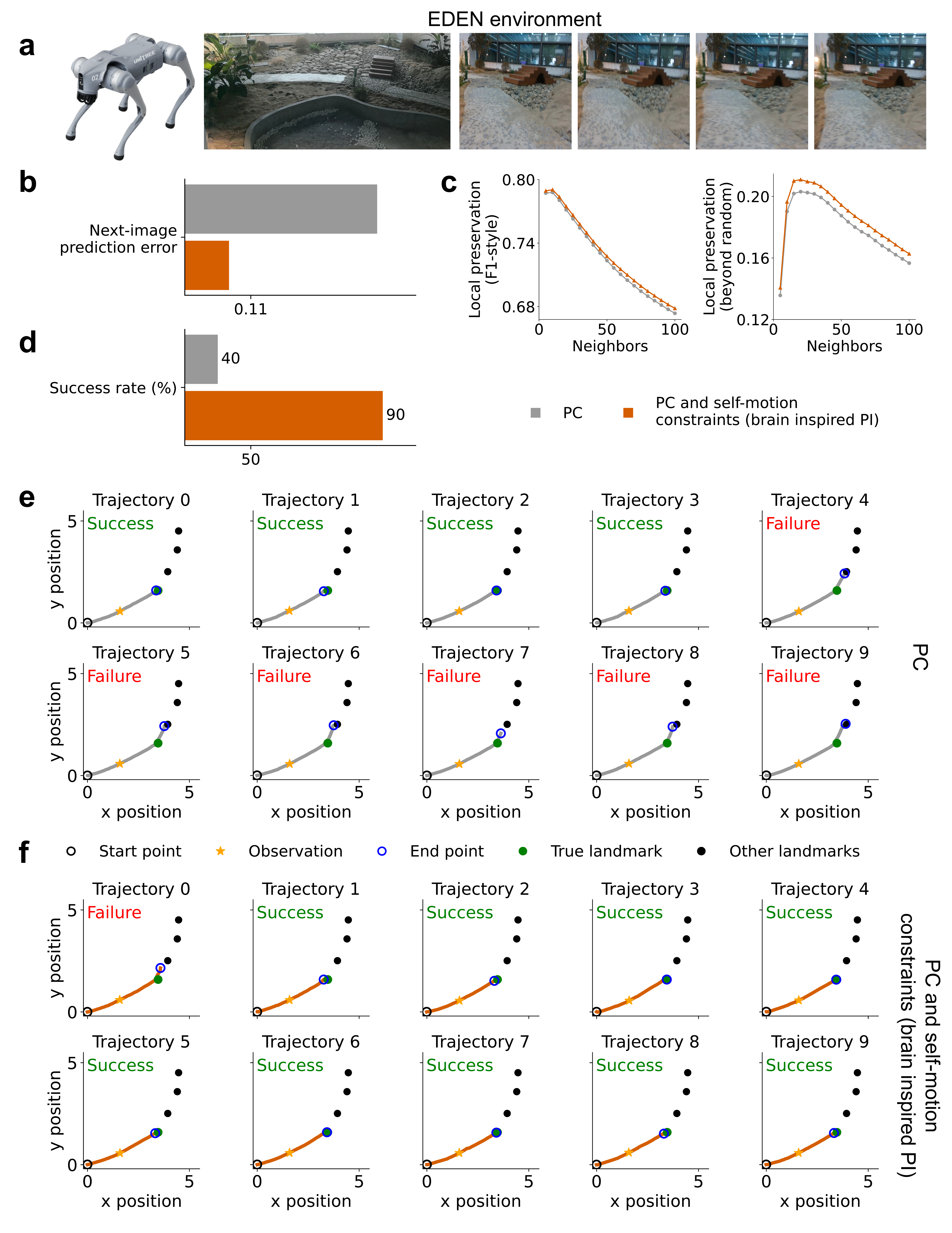}
\caption{
\textbf{Brain-inspired self-motion constraints enhance robustness of cognitive maps under real-world sensory variability.}
\textbf{a}  The real-world EDEN environment used for testing on a quadruped robot.
\textbf{b}  Offline evaluation shows that incorporating brain-inspired self-motion constraints reduces next-image prediction error in EDEN.
\textbf{c}  The constrained model yields stronger local neighborhood preservation in offline tests.
\textbf{d}  In an online landmark-identification task, the constrained model achieves a substantially higher success rate than the unconstrained baseline.
\textbf{e,f} Representative online rollouts without (e) and with (f) the self-motion constraint, illustrating that the constraint stabilizes latent state evolution and enables more reliable landmark-directed navigation.
}
\label{res4}
\end{figure}

\subsection{Self-motion constraints act as a general scaffold, with the brain-inspired prior yielding the most reliable gains}

If self-motion acts as a structural prior, its benefits should not depend on the specific implementation. 
Here, we test this claim explicitly while distinguishing general effects from implementation-dependent differences in robustness.
We compare three forms of self-motion constraint: simple velocity inputs, a rate-based PI, and the full brain-inspired PI.

Across tasks, even the weakest form of constraint proves beneficial (\textbf{Figure~\ref{res3}a}). 
In next-image prediction, all three constraints substantially reduced error, with velocity inputs, rate-based PI, and the brain-inspired PI yielding similar improvements (43.0\%, 42.1\% and 41.8\% in a highly aliased environment; 14.4\%, 13.9\% and 14.0\% in the dynamic indoor environment; and 12.7\%, 14.1\% and 14.7\% in the forest-cave-river environment). These results indicate that predictive continuity benefits broadly from self-motion information, independent of its specific form. 

The structure of the learned maps, however, shows a clearer hierarchy once geometric fidelity is considered. PI-based constraints yield larger improvements in local neighborhood preservation than velocity alone, and the brain-inspired PI consistently produces the strongest gains (\textbf{Figure~\ref{res3}b}). Velocity inputs modestly improve $\mathrm{H}(k)$ and $\mathrm{LCMC}(k)$, whereas rate-based PI produces larger gains, and the brain-inspired PI improves both metrics most substantially across all three environments. At very small neighborhood scales ($k=5$), PI-based constraints can sometimes reduce $\mathrm{LCMC}$, likely reflecting the stronger spatially meaningful structure introduced by integration, which smooths appearance-driven micro-clusters, whereas velocity inputs preserve these ultra-local features.

However, absolute-position estimation reveals a clearer dissociation between constraints (\textbf{Figure~\ref{res3}c}). Neither velocity inputs nor rate-based PI improve accuracy consistently across environments, and both increase error in some cases. By contrast, the brain-inspired PI reduced error in all environments (14.1\%, 5.1\% and 16.0\%). Visualizations of decoded position further highlight this distinction: only the brain-inspired constraint consistently sharpens global positional estimates (\textbf{Figure~\ref{res3}d}).

\textbf{Supplementary Figures~\ref{sup3} and \ref{sup4}, Tables~\ref{tab1} and \ref{tab2}} further show that these gains are not tied to a particular loss and corresponding hyperparameter. We consider three loss formulations: mean squared error (MSE), grid-augmented MSE, and Kullback–Leibler (KL)-based objectives, all yielding similarly reliable improvements, indicating that the effect arises from the structural prior itself rather than the choice of training objective.

Together, these results show that self-motion constraints serve as a generally beneficial scaffold for cognitive-map formation, despite variability across individual metrics. The brain-inspired prior provides the most consistent and comprehensive improvements.

\subsection{Brain-inspired self-motion constraints reveal superior robustness under real-world validation}

The improvements observed in simulation raise an important question: do the same benefits persist under real-world sensory conditions, where noise, environmental dynamics, and limited field-of-view pose additional challenges not captured in simulation? To assess this, we collect a new dataset in a natural indoor environment, EDEN (\textbf{Figure~\ref{res4}a; Methods}), and train two models offline: one incorporating brain-inspired self-motion constraints and one without such constraints. This comparison isolates the contribution of self-motion structure under otherwise identical training conditions in realistic settings.

The constrained model achieves lower next-image prediction error and stronger local neighborhood preservation than the unconstrained model (\textbf{Figure~\ref{res4}b and c}). Because EDEN consists of a single curved trail with relatively low perceptual ambiguity, the offline performance gap is narrower than in simulated environments. However, deploying the two pretrained models on a quadruped robot for an online landmark-identification task reveals a marked difference in robustness (see \textbf{Methods}). Across ten trials, the constrained model successfully identifies and navigates to the nearest landmark in 9 trials, whereas the unconstrained model succeeds in only 4 (\textbf{Figure~\ref{res4}d-f}). All failures involve confusing the second-closest landmark for the nearest one, an error mode plausibly driven by visual similarity and transient state instability. These results show that models equipped with brain-inspired constraints substantially reduce visual ambiguity, yielding more stable spatial estimates and more reliable physical navigation.

\section{Discussion}\label{sec3} 

Our work challenges the long-standing assumption that spatial representations in biological and artificial systems are stabilized primarily through sensory anchoring, with self-motion treated as a secondary, drift-prone update signal \cite{tolman1946studies, mcnaughton2006path, savelli2019origin, whittington2020tolman}. By embedding a motion-derived constraint into a predictive-coding framework, we demonstrate that self-motion can act as a structural prior that organizes the geometry and temporal coherence of learned cognitive maps, particularly when sensory information is unreliable or underspecified. This structural role manifests as improved local topological preservation, strengthened global positional information and enhanced predictive continuity, and emerges consistently across simulated and real-world environments. Predictive coding provides a natural testbed for structural priors, but the principle demonstrated here is not specific to PC and may extend to other representation-learning frameworks that rely on learned latent dynamics.

Here, we use the term “structural prior” to denote a constraint that limits the admissible evolution of latent states during learning, rather than an additional trainable input or regularization term. Importantly, the stabilizing effect of self-motion cannot be reduced to generic regularization or additional supervision alone. Rather than uniformly suppressing variability, the motion-derived constraint imposes trajectory-consistent structure on the latent space, selectively stabilizing representations in regimes where sensory cues alone fail to uniquely specify location. This distinction aligns with recent views that emphasize geometric and dynamical constraints on representation learning, rather than purely statistical smoothing \cite{whittington2022build, behrens2018cognitive}.

The mechanistic advantages observed in our brain-inspired path-integration module further illuminate why such structural constraints are effective under tight capacity limits. The combination of recurrent spiking dynamics, analog modulation and adaptive thresholds yields a motion prior that converges rapidly, maintains geometric fidelity and remains robust as spatial resolution increases and internal-state capacity decreases. Importantly, CA3 is not invoked as a functional model of biological path integration \cite{mcnaughton2006path, moser2017spatial}. Instead, its recurrent architecture serves as an organizational template, providing an existence proof that compact and noisy systems can nevertheless sustain precise and stable recurrent computation \cite{neunuebel2014ca3, mizuseki2012activity}. By abstracting these principles rather than reproducing biological circuitry, the model captures the computational benefits of CA3-like dynamics without overcommitting to biological specificity.

Beyond navigation, these findings point to a broader principle for representation learning. Embedding structural priors into learned latent spaces may enhance the stability of larger-scale visuomotor architectures \cite{levine2016end}, world-model agents \cite{ha2018world, wu2023daydreamer} and memory-augmented transformers \cite{dai2019transformer, bulatov2022recurrent}, where geometric consistency can improve sample efficiency and robustness. The compact spiking dynamics developed here may additionally benefit neuromorphic platforms and spiking-based state-estimation systems \cite{sun2025algorithm, sun2025exploiting, achterberg2023spatially}, where energy constraints and noise robustness are central. At the embodied level, motion-derived structural priors improve the robustness of online navigation under real-world sensory variability when deployed on a quadrupedal robot.
Rather than serving as a proof-of-concept application, the robotic experiment functions as a stress test of the proposed structural prior under real-world sensory and physical uncertainty.
Building on this observation, motion-derived structural priors may also support SLAM-free navigation and distributed exploration in more complex embodied settings \cite{wang2025multi, deng2025mne, zhao2025slam}.

Finally, while our work highlights the value of trajectory-consistent structural constraints, it also raises broader questions. Different spatial representations may act as structural priors with distinct effects on latent geometry, and determining which forms best stabilize learned world models remains an open problem. Our exploratory experiments suggest that directly imposing grid-like templates can conflict with evolving sensory manifolds, whereas path-integration-based constraints integrate more naturally with predictive-coding dynamics. Understanding how such priors interact with sensory evidence, and how they shape representation learning under uncertainty, will be essential for advancing theories of spatial cognition and for designing autonomous agents capable of maintaining coherent internal structure in uncertain environments.

\section{Methods}\label{sec4}

\subsection{Predictive coding}

The predictive coding processes a visual sequence $\{x_1, x_2, \ldots, x_t\}$ and serves two purposes: (1) predicting the next visual observation $\hat{x}_{t+1}$ to support map construction \cite{gornet2024automated}, and (2) predicting a relative position representation $\hat{p}_{t+1}$ through an auxiliary position-decoding head that interfaces with the PI-based geometric constraint. At each time step $t$, the visual input $x_t$ is encoded by a visual encoder $E_{\phi}$ into a latent representation $z_t$. The latent sequence $z_{1:t}$ is then processed by a sequence model $S_{\phi}$ with recurrent latent state $h_t$ to produce the next latent state $h_{t+1}$. Two decoder heads, $D_{\text{img},\phi}$ and $D_{\text{pos},\phi}$, map $h_{t+1}$ to the next-step visual prediction $\hat{x}_{t+1}$ and the relative position prediction $\hat{p}_{t+1}$, respectively. The formulation is:
\begin{subequations}
\begin{equation}
z_t = E_{\phi}(x_t),
\label{eq:enc}
\end{equation}

\begin{equation}
h_{t+1} = S_{\phi}(h_t, z_{1:t}),
\label{eq:seq}
\end{equation}

\begin{equation}
\hat{x}_{t+1} = D_{\text{img},\phi}(h_{t+1}), 
\quad 
\hat{p}_{t+1} = D_{\text{pos},\phi}(h_{t+1}).
\label{eq:dec}
\end{equation}
\end{subequations}
The position decoder $D_{\text{pos},\phi}$ is implemented as a multilayer perceptron (MLP) preceded by a convolutional layer for dimensionality reduction. Its output dimensionality is chosen to match the resolution of the place-cell representation. The remaining components follow \cite{gornet2024automated}: $E_{\phi}$ uses a ResNet-18 backbone \cite{he2016deep}, $D_{\text{img},\phi}$ mirrors this architecture with transposed convolutions, and $S_{\phi}$ is implemented as a multi-headed self-attention model \cite{vaswani2017attention}.

\subsection{Brain-inspired PI}

The PI module supplies the self-motion prior used to constrain the predictive coding model. Given the current velocity input $v_t$, the PI predicts the next-step position representation $p_{t+1}$, which serves as the target for the PC’s auxiliary position prediction. The PI is pretrained on simulated rodent trajectories \cite{raudies2012modeling, banino2018vector}, and its parameters are fixed during cognitive map construction, denoted $\theta^*$.

At each time step $t$, the motion input $v_t$ is processed by a sequence model $S_{\theta^*}$ with recurrent state $g_t$ to produce the next latent state $g_{t+1}$. This state is then passed to a position decoder $D_{\theta^*}$ to produce the position representation $p_{t+1}$. The initial position of each sequence is set to $\text{pos}_1 = (0, 0)$. This position is first encoded by a place-cell encoder $E_{\text{pc}}$ and then embedded by $E_{\theta^*}$ to initialize the recurrent state $g_1$ for $S_{\theta^*}$. The computations are:
\begin{subequations}
\begin{equation}
g_{t+1} = S_{\theta^*}(g_t, v_t),
\qquad
g_1 = E_{\theta^*}(E_{\text{pc}}(\text{pos}_1)),
\qquad
\text{pos}_1 = (0, 0),
\label{eq:seq_pi}
\end{equation}

\begin{equation}
p_{t+1} = D_{\theta^*}(g_{t+1}).
\label{eq:dec_pi}
\end{equation}
\end{subequations}
The place-cell encoder $E_{\text{pc}}$ adopts a center-surround place coding model \cite{banino2018vector}, and both $E_{\theta^*}$ and $D_{\theta^*}$ are implemented as fully connected layers.

To obtain a more precise yet capacity-efficient source of self-motion information, we construct a brain-inspired PI module whose recurrent core $S_{\theta}$ is based on recurrent spiking dynamics with analog modulation and adaptive thresholds, consistent with CA3 physiology \cite{alle2006combined, sasaki2011action, zbili2019past}. The recurrent dynamics are defined as:

\begin{subequations}
\begin{equation}
V_{t+1} = \tau_s V_t + (1-\tau_s)(W_{\text{in}} v_t + W_{\text{rec}} g_t),
\label{eq:amrsnn1}
\end{equation}

\begin{equation}
s_{t+1} =
\begin{cases}
a_{t+1}, & \text{if } V_{t+1} \ge V_{\text{thr}}^{\theta},\\
0, & \text{otherwise},
\end{cases}
\label{eq:amrsnn2}
\end{equation}

\begin{equation}
a_{t+1} = \text{clip}(V_{t+1},0,1) =
\begin{cases}
0, & V_{t+1} < 0,\\
V_{t+1}, & 0 \le V_{t+1} \le 1,\\
1, & V_{t+1} > 1,
\end{cases}
\label{eq:amrsnn3}
\end{equation}

\begin{equation}
g_{t+1} = \alpha g_t + (1-\alpha)s_{t+1}.
\label{eq:amrsnn4}
\end{equation}
\end{subequations}
Here, $V_t$ denotes the membrane potential, $W_{\text{in}}$ and $W_{\text{rec}}$ are the input and recurrent weight matrices of $S_{\theta}$, and $\tau_s$ is the somatic decay factor. A spike is emitted whenever $V_{t+1}$ exceeds the learnable threshold $V_{\text{thr}}^{\theta}$, with its analog amplitude $a_{t+1}$ given by the clipped membrane potential. The latent position state $g_{t+1}$ is updated via leaky integration of the analog spike $s_{t+1}$, controlled by the decay factor $\alpha$.

\subsection{Formulation of self-motion constraints}

The self-motion constraints are implemented through a geometric loss term $L_{\text{geo}}$, which aligns the position representations produced by the PC module with those generated by the PI module. Among the several formulations we evaluate in the main text, we present here the KL-based version used in the primary model. In addition, the PC module is trained with an MSE reconstruction loss $L_{\text{img}}$ that evaluates next-image prediction. The geometric constraint $L_{\text{geo}}$ is weighted by a hyperparameter $\lambda$, which controls the strength of the constraint, and added to the reconstruction loss to form the final objective. The training losses are defined as:
\begin{subequations}
\begin{equation}
L_\text{geo}(\phi) = \sum_{t=2}^T \text{KL}(\hat{p}_t \,\|\, p_t), \quad L_\text{img}(\phi) = \sum_{t=2}^T \text{MSE}(\hat{x}_t, x_t),
\label{eq:loss_gi}
\end{equation}

\begin{equation}
L_\text{total}(\phi) = L_\text{img}(\phi) + \lambda L_{\text{geo}}(\phi).
\label{eq:loss_total}
\end{equation}
\end{subequations}
Since $t = 1$ corresponds to the initialization step, both the image-prediction error and the position-prediction error are computed from $t = 2$ onward, where $T$ denotes the sequence length. Because the PI module parameters are frozen, only the PC module parameters $\phi$ are updated during training. At inference time, the PC module’s hidden state $h_t$ is used to assess the model’s representation of the environment and to perform downstream tasks. Consequently, the self-motion constraints incur no additional parameter cost at inference.

\subsection{Environments}
\subsubsection{Simulation environments}
We evaluate our models in three simulated environments: a highly-aliased environment, a dynamic-indoor environment, and a forest-cave-river environment. All environments are implemented using the Malmo platform \cite{johnson2016malmo}, which enables controlled visual and geometric configuration within Minecraft. Following prior work \cite{gornet2024automated}, the agent navigates between randomly sampled waypoints. In contrast to previous settings, we additionally record ego-motion velocity at every time step.

To assess the ability to disambiguate visually similar locations, we construct a highly-aliased environment (\textbf{Figure~\ref{res1}a}). It contains two visually identical rooms that partially overlap, with two doors in the overlapping region leading to different rooms. Each room includes four sparsely placed colored flags as landmarks, while most adjacent locations share nearly indistinguishable appearances, creating strong visual aliasing.

To evaluate robustness to visual dynamics, we design a dynamic indoor environment (\textbf{Figure~\ref{res1}a}). This environment includes multiple glass windows and a glass ceiling, allowing illumination to vary continuously with time-of-day and weather conditions. A moving minecart introduces additional temporal appearance changes.

To examine performance in naturalistic outdoor settings, we adopt the forest-cave-river environment from \cite{gornet2024automated} (\textbf{Figure~\ref{res1}a}). It contains three prominent visual regions: a large cave that serves as a global landmark, a densely vegetated forest, and a river segment connected by a bridge.

\subsubsection{Real-world environment}
To assess spatial cognition under real-world deployment conditions, we evaluate our models in a physical environment referred to as EDEN (\textbf{Figure~\ref{res4}a}). The environment includes five visually distinctive landmarks: a curved pathway, a bridge, a pond, gravel piles, and vegetation. For data collection, a quadruped robot is teleoperated along the curved pathway while first-person visual observations and ego-motion velocities are recorded.

\subsection{Training and evaluation of the cognitive map construction}
We describe here the procedures used to construct and evaluate the cognitive map. All configurations and hyperparameters are listed in \textbf{Supplementary Table~\ref{tab3}}.

\subsubsection{Training}
In all three simulated environments, the models are trained from scratch under identical settings: an input sequence length of 10, a batch size of 16, and 200 training epochs. We use stochastic gradient descent with a learning rate of 0.1, a weight decay of $5 \times 10^{-6}$, and the OneCycle learning-rate schedule. During cognitive-map construction, the PI module remains frozen and only the PC module is updated.

For the real-world environment, the visual encoder is initialized with ImageNet-pretrained weights due to the limited amount of real data, while all downstream components are trained from scratch. The PI module is likewise kept frozen. We use the same input sequence length (10) and batch size (16), and train for 50 epochs with the Adam optimizer (learning rate $1 \times 10^{-3}$) and the OneCycle learning-rate schedule. 

\subsubsection{Metrics}
To assess whether the latent space preserves the local neighborhood structure of the underlying spatial domain, we use standard manifold-learning metrics, Trustworthiness $\mathrm{T}(k)$ and Continuity $\mathrm{C}(k)$, computed at neighborhood size $k$. Trustworthiness evaluates the extent to which the latent space avoids introducing false neighbors (precision-like), whereas Continuity evaluates the extent to which it retains true neighbors (recall-like). Their harmonic mean,
\begin{subequations}
\begin{equation}
    \text{H}(k) = \frac{2 \, \text{T}(k) \text{C}(k)}{\text{T}(k) + \text{C}(k)},
\label{eq:Hk}
\end{equation}
provides an F1-style summary of local structure preservation.
The Trustworthiness metric is defined as:
\begin{equation}
    \text{T}(k) = 1 - \frac{2}{n k (2n - 3k - 1)}
    \sum_{i=1}^{n} \sum_{j \in U_i(k)} (r_X(i, j) - k),
    \quad U_i(k) = N_k^Z(i) \setminus N_k^X(i),
\label{eq:Tk}
\end{equation}
and Continuity is defined analogously:
\begin{equation}
    \text{C}(k) = 1 - \frac{2}{n k (2n - 3k - 1)}
    \sum_{i=1}^{n} \sum_{j \in V_i(k)} (r_Z(i, j) - k),
    \quad V_i(k) = N_k^X(i) \setminus N_k^Z(i).
\label{eq:Ck}
\end{equation}
\end{subequations}
Here, $X$ denotes the input space and $Z$ denotes the latent space; $N_k^X(i)$ and $N_k^Z(i)$ are the $k$-nearest neighbors of sample $i$ in $X$ and $Z$, respectively, and $r_X(i,j)$ and $r_Z(i,j)$ denote the rank of sample $j$ in the corresponding spaces. Both $\mathrm{T}(k)$ and $\mathrm{C}(k)$ are normalized to $[0,1]$, with higher values indicating better preservation of local neighborhood structure. In experiments, we report $\mathrm{H}(k)$ curves as the primary summary measure.

To complement $\mathrm{H}(k)$, we also report the local continuity meta-criterion $\mathrm{LCMC}(k)$, which quantifies the overlap between $k$-nearest neighbor sets in $X$ and $Z$ while subtracting the expected overlap under random mapping:
\begin{equation}
    \text{LCMC}(k) = \frac{1}{n k} \sum_{i=1}^{n} \big| N_k^X(i) \cap N_k^Z(i) \big| \;-\; \frac{k}{n - 1}.
\label{eq:LCMC}
\end{equation}
The term $\tfrac{k}{n-1}$ corresponds to the expected overlap under random mapping; thus, $\mathrm{LCMC}(k)=0$ denotes random-level performance, positive values indicate better-than-random neighborhood preservation, and negative values indicate worse-than-random performance.

Whereas $\mathrm{H}(k)$ captures precision-recall trade-offs in local neighborhoods, $\mathrm{LCMC}(k)$ measures the degree to which locality in the latent space exceeds the random baseline. Together, these metrics provide a comprehensive characterization of local geometric and topological fidelity.

\subsubsection{Downstream tasks}

To evaluate absolute-position prediction, we follow \cite{gornet2024automated} and train an auxiliary neural network that maps the latent state $h_t$ to a 2D absolute position. The network consists of a convolution-pooling block followed by two linear layers.

Real-world online performance is assessed using a landmark-identification task. Four landmarks are predefined, and the latent states $h_t$ corresponding to these landmarks in the training trajectories are stored in a memory bank as reference embeddings. During testing, the latent state extracted from the query observation is compared against the memory bank, and the nearest embedding is used to identify the corresponding landmark. The robot then follows the associated waypoint sequence to navigate toward the predicted landmark, providing real-world validation of the learned spatial representations.

\subsection{Pretraining configuration of the PI module}
In this section, we describe the configuration used to pretrain the PI module. All architectural and training hyperparameters are summarized in \textbf{Supplementary Table~\ref{tab4}}.

\subsubsection{Simulated rodent trajectories}
The PI module is pretrained on simulated rodent trajectories \cite{raudies2012modeling, banino2018vector}, generated as random walks within a rectangular arena. Trajectories are sampled at a fixed time step of 0.1\,s. Walking speed is drawn from a Rayleigh distribution, and turning angles are sampled from a Gaussian distribution. When the agent approaches a wall, its speed is reduced and its heading is redirected away from the boundary. We use a $1.2 \times 1.2$ arena, a Rayleigh scale parameter of $0.18$ for speed sampling, and a Gaussian distribution with mean $0$ and standard deviation $2.304$ for angular velocity. Each trajectory consists of 20 time steps.

\subsubsection{Training}

The PI module is trained to reconstruct ground-truth place-cell activations. 
Let $y$ denote the target place-cell activity and $\hat{y}$ the PI prediction. 
Training minimizes the cross-entropy between $y$ and $\hat{y}$, together with an L2 penalty on the recurrent weights:
\begin{equation}
L_{\text{PI}}
= \text{CE}(\hat{y}, y)
+ \lambda_{\text{PI}} \,\| W_{\text{rec}} \|_2^2 .
\label{eq:loss_pi}
\end{equation}
Here, $W_{\text{rec}}$ denotes the recurrent weight matrix of the sequence model $S_{\theta}$ and $\lambda_{\text{PI}}$ is the weight decay coefficient. 
Training uses the Adam optimizer with a batch size of 200.

We evaluate five PI variants: rate-based PI, spike-based PI, spike-based PI with learnable thresholds, spike-based PI with analog modulation and the full brain-inspired PI. 
For all spike-based models, we use a learning rate of $1 \times 10^{-3}$ and weight decay of $1 \times 10^{-3}$. For the rate-based model, the learning rate is reduced to $1 \times 10^{-4}$ and the weight decay to $1 \times 10^{-4}$, as the spike-based hyperparameters lead to non-convergence.

\subsubsection{Place coding}
We adopt a center-surround place-cell encoding scheme \cite{banino2018vector} to map each 2D position to an $n_p$-dimensional activity pattern. Place cells are uniformly tiled across the simulated environment, and the response of the $i$-th cell is computed as the difference of two Gaussian kernels:
\begin{equation}
\begin{aligned}
y_i(X_{\text{pos}})
&=
\frac{\exp\!\big(-\|X_{\text{pos}} - c_i\|^2 / (2\sigma_1^2)\big)}
     {\sum_{j=1}^{n_p} \exp\!\big(-\|X_{\text{pos}} - c_j\|^2 / (2\sigma_1^2)\big)}
\\
&\quad -
\frac{\exp\!\big(-\|X_{\text{pos}} - c_i\|^2 / (2\sigma_2^2)\big)}
     {\sum_{j=1}^{n_p} \exp\!\big(-\|X_{\text{pos}} - c_j\|^2 / (2\sigma_2^2)\big)} .
\end{aligned}
\label{eq:place_res}
\end{equation}
Here, $X_{\text{pos}} \in \mathbb{R}^2$ denotes the agent’s position, $c_i$ the center of the $i$-th place cell, and $\sigma_1$ and $\sigma_2$ the widths of the excitatory and inhibitory kernels ($\sigma_2 > \sigma_1$). We use $\sigma_1 = 0.06$ and $\sigma_2 = 0.12$ in all experiments. The resulting activity vector is normalized to lie in $[0,1]$ and sums up to one.

\bibliography{sn-bibliography}
\clearpage
\appendix

\clearpage
\appendix

\makeatletter
\g@addto@macro\appendix{
  \renewcommand{\thefigure}{S\arabic{figure}}
  \renewcommand{\thetable}{S\arabic{table}}
}
\makeatother

\section*{Supplementary Information}
\setcounter{figure}{0}
\setcounter{table}{0}

\section*{Supplementary Figures}

\begin{figure}[h]
\centering
\includegraphics[width=\linewidth]{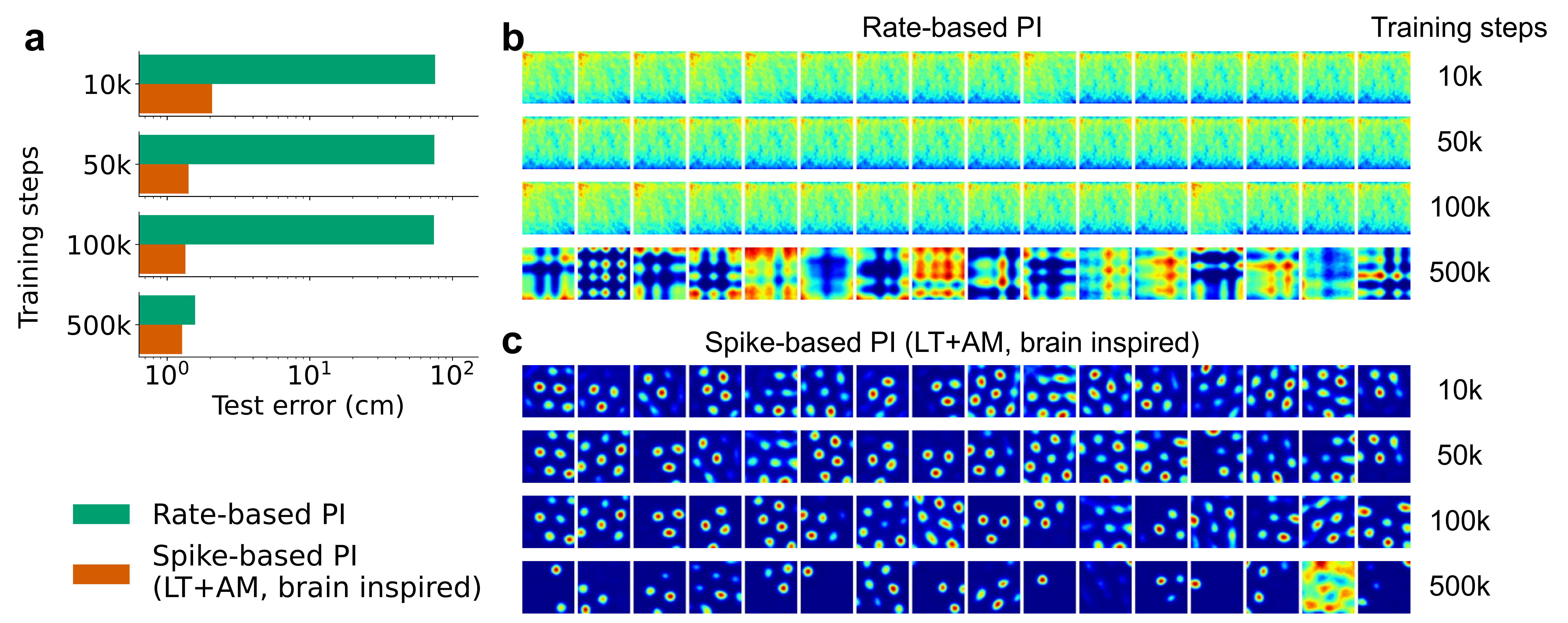}
\caption{
\textbf{Brain-inspired path integration converges rapidly and supports earlier emergence of structured spatial codes (256 hidden units, 3600 place cells).}
\textbf{a}  Training dynamics show that the brain-inspired PI reaches a low-error regime by approximately 50k steps, whereas the rate-based PI requires nearly 500k steps to achieve comparable error levels.
\textbf{b}  At 500k training steps, the rate-based PI exhibits weak and spatially diffuse square-like periodicity.
\textbf{c}  By contrast, at 50k steps the brain-inspired PI already expresses clear hexagonal grid-like spatial patterns. 
}
\label{sup1}
\end{figure}

\clearpage
\begin{figure}[h]
\centering
\includegraphics[width=\linewidth]{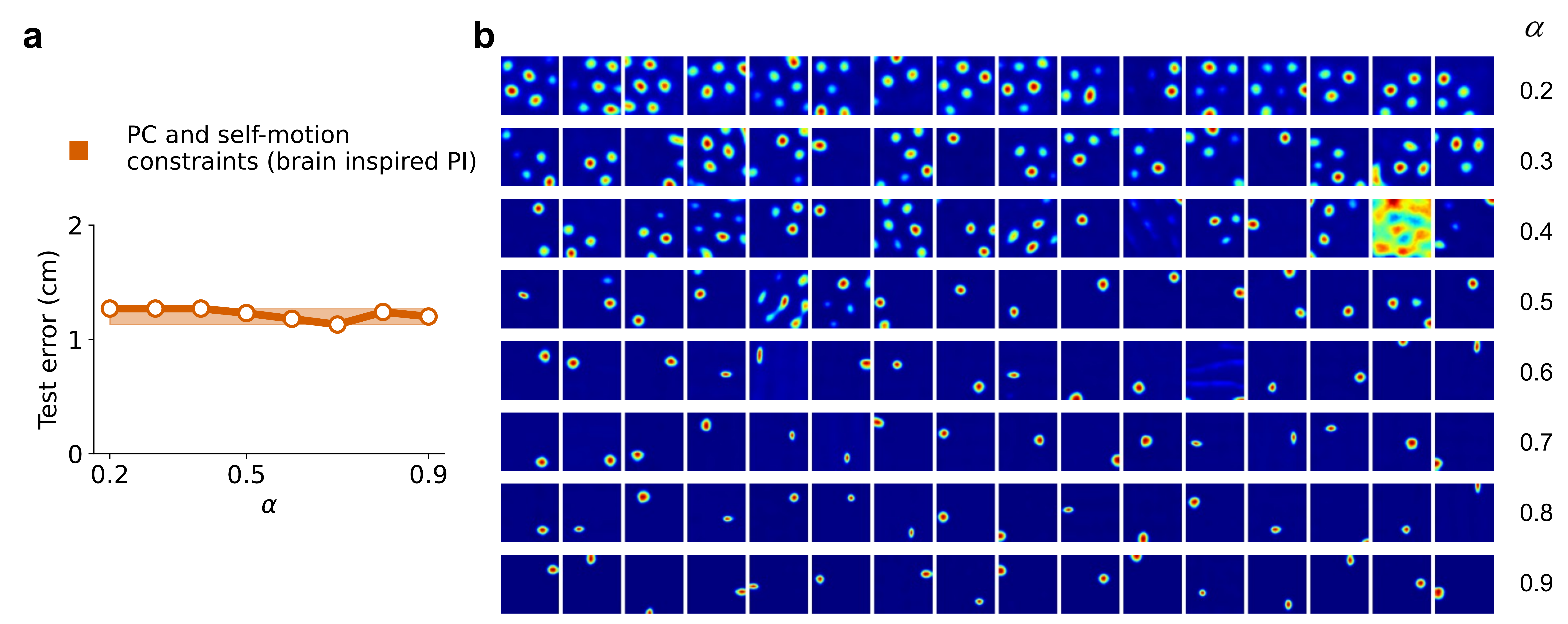}
\caption{
\textbf{The brain-inspired path integration remains robust across latent-state integration factor $\alpha$ and exhibits a smooth transition in spatial response structure (256 hidden units, 3600 place cells).}
\textbf{a}  Across a broad range of $\alpha$ values, the brain-inspired PI maintains consistently low integration error, indicating that performance is not dependent on precise hyperparameter tuning.
\textbf{b}  Spatial responses vary smoothly with $\alpha$, transitioning from pronounced grid-like periodicity toward increasingly localized, place-like patterns as $\alpha$ increases.
}
\label{sup2}
\end{figure}

\clearpage
\begin{figure}[h]
\centering
\includegraphics[width=0.98\linewidth]{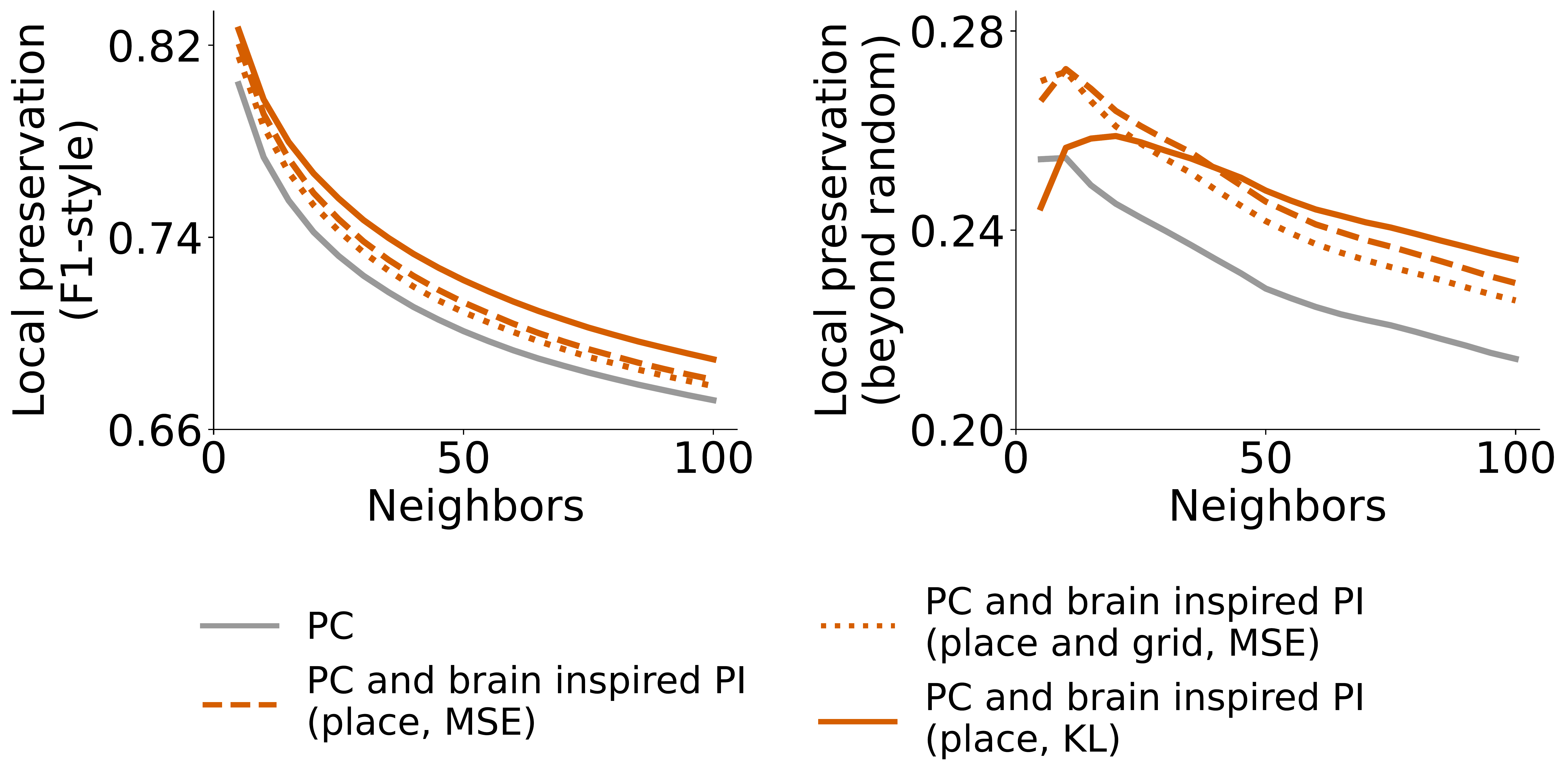}
\caption{
\textbf{The brain-inspired path integration constraint provides consistent improvements in local neighborhood preservation across three alternative geometric-constraint formulations.} 
The evaluated formulations include (1) aligning the PC-decoded positional representation with the PI-derived place-cell representation using an MSE objective, (2) augmenting this MSE objective with latent grid-alignment, and (3) replacing the MSE objective with a KL-divergence objective. 
Neighborhood structure is evaluated in a highly aliased environment using $\mathrm{H}(k)$ (F1-like) and $\mathrm{LCMC}(k)$ (beyond a random-overlap baseline). 
All metrics are averaged over the final 10 training epochs.
}
\label{sup3}
\end{figure}

\clearpage
\begin{figure}[h]
\centering
\includegraphics[width=0.98\linewidth]{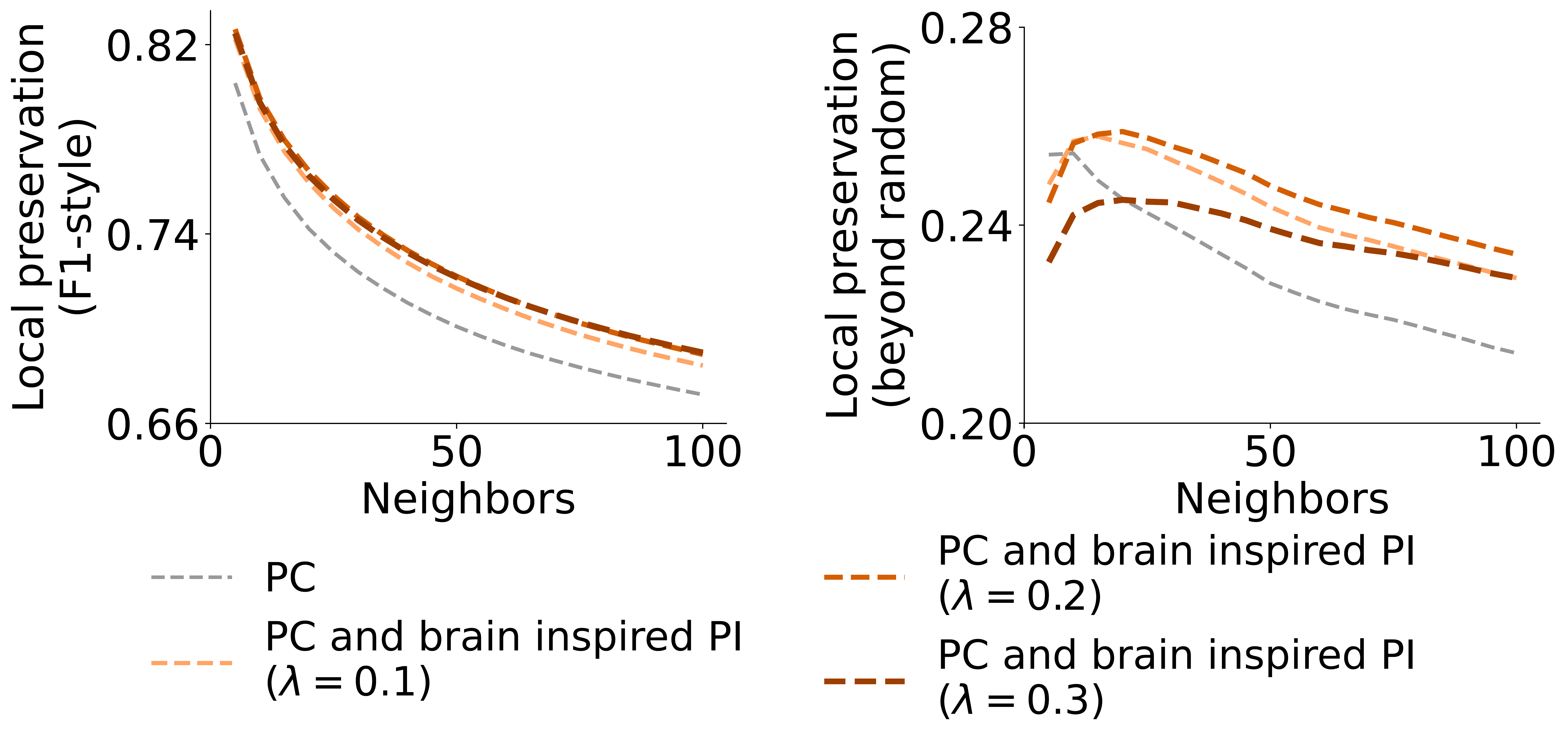}
\caption{
\textbf{The brain-inspired constraint yields consistent gains across a range of constraint strengths.}
Here, $\lambda$ controls the strength of the geometric constraint in the training objective. Three settings ($\lambda$ = 0.1, 0.2, 0.3) are evaluated in a highly aliased environment, with neighborhood preservation quantified using $\mathrm{H}(k)$ (harmonic mean of Trustworthiness and Continuity) and $\mathrm{LCMC}(k)$ (beyond a random-overlap baseline). 
All metrics are averaged over the final 10 training epochs.
}
\label{sup4}
\end{figure}

\clearpage
\section*{Supplementary Tables}
\begin{table}[h]
\caption{
The brain-inspired PI yields generally consistent performance gains in next-image prediction and absolute-position estimation under the same three geometric-constraint formulations. 
The formulations respectively use (1) an MSE objective to align PC-decoded and PI-derived representations, (2) this MSE objective with additional latent grid-alignment, or (3) a KL-divergence objective in place of MSE, all evaluated in a highly aliased environment. 
Next-image errors are averaged over the final 10 epochs, and absolute-position errors over the final 20 epochs.
}
\label{tab1}
\begin{tabular}{@{}cccc@{}}
\toprule
\multicolumn{4}{c}{Next-image prediction error} \\ \midrule
PC & \begin{tabular}[c]{@{}c@{}}PC and brain inspired PI\\ (place, MSE)\end{tabular} & \begin{tabular}[c]{@{}c@{}}PC and brain inspired PI\\ (place and grid, MSE)\end{tabular} & \begin{tabular}[c]{@{}c@{}}PC and brain inspired PI\\ (place, KL)\end{tabular} \\
0.0661 & 0.0373 & 0.0373 & 0.0385 \\ \midrule
\multicolumn{4}{c}{Absolute-position prediction error} \\ \midrule
PC & \begin{tabular}[c]{@{}c@{}}PC and brain inspired PI\\ (place, MSE)\end{tabular} & \begin{tabular}[c]{@{}c@{}}PC and brain inspired PI\\ (place and grid, MSE)\end{tabular} & \begin{tabular}[c]{@{}c@{}}PC and brain inspired PI\\ (place, KL)\end{tabular} \\
0.0497 & 0.0512 & 0.0429 & 0.0427 \\ \bottomrule
\end{tabular}
\end{table}
We ask whether the gains of the brain-inspired PI depend on the precise constraint formulation or remain stable across alternatives. To investigate this question, we compare three formulations of the constraint in the highly aliased environment. 
The first formulation aligns the output of the PC position decoder with the PI-derived place-cell representation using a MSE objective. The second formulation augments this objective by additionally aligning latent grid-like responses. The third formulation replaces the MSE objective with a KL objective. All three formulations yield robust improvements relative to models trained without self-motion constraints. In next-image prediction, error decreases by 41.8\% to 43.6\%. In absolute-position estimation, the formulations that incorporate grid alignment or the KL objective reduce error by 13.7\% to 14.1\%. Local neighborhood preservation also improves reliably across formulations, with increases of 1.2\% to 2.9\% in $\mathrm{H}(k)$ and 5.9\% to 7.3\% in $\mathrm{LCMC}(k)$ on average (\textbf{Supplementary Figure~\ref{sup3}}).

\clearpage
\begin{table}[h]
\caption{
The brain-inspired constraint yields generally consistent gains in next-image prediction and absolute-position estimation across three choices of $\lambda$ in the KL-based objective. 
We evaluate $\lambda \in \{0.1, 0.2, 0.3\}$, where $\lambda$ modulates the influence of the constraint, in the same highly aliased environment. 
Next-image errors are averaged over the final 10 epochs, and absolute-position errors over the final 20 epochs.
}
\label{tab2}
\begin{tabular}{@{}cccc@{}}
\toprule
\multicolumn{4}{c}{Next-image prediction error} \\ \midrule
PC & \begin{tabular}[c]{@{}c@{}}PC and brain inspired PI\\ ($\lambda=0.1$)\end{tabular} & \begin{tabular}[c]{@{}c@{}}PC and brain inspired PI\\ ($\lambda=0.2$)\end{tabular} & \begin{tabular}[c]{@{}c@{}}PC and brain inspired PI\\ ($\lambda=0.3$)\end{tabular} \\
0.0661 & 0.0376 & 0.0385 & 0.0385 \\ \midrule
\multicolumn{4}{c}{Absolute-position prediction error} \\ \midrule
PC & \begin{tabular}[c]{@{}c@{}}PC and brain inspired PI\\ ($\lambda=0.1$)\end{tabular} & \begin{tabular}[c]{@{}c@{}}PC and brain inspired PI\\ ($\lambda=0.2$)\end{tabular} & \begin{tabular}[c]{@{}c@{}}PC and brain inspired PI\\ ($\lambda=0.3$)\end{tabular} \\
0.0497 & 0.0453 & 0.0427 & 0.0480 \\ \bottomrule
\end{tabular}
\end{table}
We investigate how the strength of the brain-inspired constraint influences performance by evaluating three strengths of the KL-based objective (0.1, 0.2 and 0.3) in the highly aliased environment. Across all strengths, the brain-inspired constraint again produces consistent improvements. In next-image prediction, errors decrease by 41.8\% to 43.1\%. In absolute-position estimation, errors decrease by 3.4\% to 14.1\%. Local neighborhood preservation follows the same trend, with $\mathrm{H}(k)$ increasing by 2.3\% to 2.9\% and $\mathrm{LCMC}(k)$ increasing by 3.0\% to 6.0\% on average (\textbf{Supplementary Figure~\ref{sup4}}).

\clearpage
\begin{table}[h]
\caption{Architecture configuration and hyperparameter settings for cognitive-map construction using the pretrained PI module.}
\label{tab3}
\begin{tabular}{@{}lll@{}}
\toprule
\textbf{Category} & \textbf{Parameter} & \textbf{Setting} \\ \midrule
\multirow{4}{*}{PC module} & Visual encoder $E_\phi$ & ResNet-18 \cite{gornet2024automated} \\
& Sequence model $S_\phi$ & Multi-head self-attention \cite{gornet2024automated} \\
& Image decoder $D_{\text{img},\phi}$ & ResNet-18 \cite{gornet2024automated} \\
& Position decoder $D_{\text{pos},\phi}$ & \begin{tabular}[c]{@{}l@{}}Conv \\ $\rightarrow$ max pool \\ $\rightarrow$ FC (256-d)\\ $\rightarrow$ FC (3600-d)\end{tabular} \\ \midrule
\multirow{5}{*}{PI module} & Sequence model $S_\theta$ & \begin{tabular}[c]{@{}l@{}}Rate-based variant;\\ Brain-inspired variant.\end{tabular} \\
& Initial-state encoder $E_{\theta}$ & FC to initial latent state \\
& Position decoder $D_{\theta}$ & FC to place-cell code \\
& Latent size $n_g$ & 256 \\
& Place cell size $n_p$ & 3600 \\ \midrule
\multirow{7}{*}{Training (simulation)} & Trainable components & \begin{tabular}[c]{@{}l@{}}PC module (trainable), \\ PI module (frozen)\end{tabular} \\
& Input sequence length & 10 \\
& Batch size / epochs & 16 / 200 \\
& Optimizer / learning rate & SGD / 0.1 \\
& Weight decay & $5 \times 10^{-6}$ \\
& Learning rate schedule & OneCycle \\
& Objective & $L_{\text{total}} = L_{\text{img}} + \lambda \, L_{\text{geo}}$ \\ \midrule
\multirow{6}{*}{Training (real-world)} & Trainable components & \begin{tabular}[c]{@{}l@{}}PC module (ImageNet-pretrained $E_\phi$, \\ finetune downstream),\\ PI module (frozen)\end{tabular} \\
& Input sequence length & 10 \\
& Batch size / epochs & 16 / 50 \\
& Optimizer / learning rate & Adam / $1\times 10^{-3}$ \\
& Learning rate schedule & OneCycle \\
& Objective & Same as simulation \\ 
\bottomrule
\end{tabular}
\end{table}

\clearpage
\begin{table}[h]
\caption{Architecture configuration and hyperparameter settings for pretraining the PI module on simulated trajectories.}
\label{tab4}
\begin{tabular}{@{}lll@{}}
\toprule
\textbf{Category} & \textbf{Parameter} & \textbf{Setting} \\ \midrule
\multirow{5}{*}{Simulated trajectory} & Arena size & $1.2 \times 1.2$ (square enclosure) \\
& Time step & $0.1\,\mathrm{s}$ \\
& Speed sampling & Rayleigh distribution (scale $0.18$) \\
& Angular velocity & Gaussian distribution (mean $0$, std. $2.304$) \\
& Trajectory length & 20 time steps \\ \midrule
\multirow{5}{*}{Architecture} & Sequence model $S_{\theta}$ & \begin{tabular}[c]{@{}l@{}}Rate-based variant;\\ spike-based variant;\\ spike-based variant with learnable threshold;\\ spike-based variant with analog modulation;\\ brain-inspired variant.\end{tabular} \\
& Initial-state encoder $E_{\theta}$ & FC to initial latent state \\
& Position decoder $D_{\theta}$ & FC to place-cell code \\
& Latent size $n_g$ & $\{256, 128\}$ (default 256) \\
& Place cell size $n_p$ & $\{400, 900, 1600, 3600\}$ (default 3600) \\ \midrule
\multirow{6}{*}{Training} & Objective & $\text{CE}(\hat{p},p) + \lambda_{\text{PI}} \|W_{\text{rec}}\|_2^2$ \\
& Optimizer & Adam \\
& Weight decay & \begin{tabular}[c]{@{}l@{}}$1\times 10^{-3}$ (spike-based variants);\\ $1\times 10^{-4}$ (rate-based variants)\end{tabular} \\
& Learning rate & \begin{tabular}[c]{@{}l@{}}$1\times 10^{-3}$ (spike-based variants);\\ $1\times 10^{-4}$ (rate-based variants)\end{tabular} \\
& Batch size & $200$ \\
& Training iterations & $\{10k, 50k, 100k, 500k\}$ (default 500k) \\ \midrule
\multirow{5}{*}{Place coding $E_{\text{pc}}$} & Model & Center-surround place coding \cite{banino2018vector} \\
& Cell placement & Uniformly distributed over the arena \\
& Excitatory width & $\sigma_1 = 0.06$ \\
& Inhibitory width & $\sigma_2 = 0.12$ \\
& Normalization & Clipped to $[0,1]$ and L1-normalized \\ 
\bottomrule
\end{tabular}
\end{table}

\end{document}